\documentclass{article}
\usepackage{fleqn}
\usepackage{epsfig}
\usepackage{amsmath,amssymb}

\begin{document}

\begin{center}
{\bf\Large The complete relativistic kinetic model of symmetry
violation
 in isotopic expanding plasma. III. Specific entropy calculation.}\\[12pt]
Yu.G.Ignat'ev, K.Alsmadi\\
Kazan State Pedagogical University,\\ Mezhlauk str., 1, Kazan
420021, Russia
\end{center}

\begin{abstract}
A complete model of baryon production in an expanding,
primordially symmetric hot Universe is constructed in the
framework of general-relativistic kinetic theory. In this model
specific model for a baryon is calculated and graphs of the value
dependence are constructed.\end{abstract}

\section{Transformation to the dimentionless variables}
In the previous papers of the authors \cite{Yu_barI},
\cite{Yu_barII} in terms of the developed relativistic kinetic
theory of baryons production\footnote{As well as the other similar
particles arising as a result of thermodynamic equilibrium
violation and spontaneous $CP$ invariance violation.} an
expression for final concentration of baryons in hot Universe was
obtained (23)\cite{Yu_barII}, using which along with the dense
entropy equation (26)\cite{Yu_barII}, we get the sought-for
expression for final dense of specific entropy accounted for one
baryon
\begin{equation}\label{ds_b}
\delta_S\!=\!\frac{N_b(\infty)}{S}=\frac{30\Delta
r\mathcal{N}_X}{\pi^4\mathcal{T}^3\mathcal{N}}\int\limits_0^\infty
\exp\left(\! -\int\limits_t^\infty \Psi(t')dt'\right)G(t)dt,
\end{equation}
where
\begin{equation}\label{III.127}
\Psi(t)=\frac{2N_X}{\pi^2 \mathcal{T}^3}\int\limits_0^\infty
\mathbb{P}^2\dot{\Phi}f_0 \beta_0 d\mathbb{P},
\end{equation}
\begin{equation}\label{III.128}
G(t)=\frac{1}{\pi^2}\int\limits_0^\infty \mathbb{P}^2
\dot{\Phi}\delta f d \mathbb{P};
\end{equation}
derivatives by time are denoted by a dot $t$, $\delta f$ - is a deviation of the
boson distribution function from
equilibrium in a symmetric plasma $\lambda=0$
\begin{equation}\label{III.121}%(II.16)
\delta f(\mathbb{P},t)=- e^{-\Phi(\mathbb{P},t)}\int\limits_0^t
e^{\Phi(\mathbb{P},t')}\dot{f}_0(0;\mathbb{P},t')dt',
\end{equation}
and the following designation is used
\begin{equation}\label{III.122}%(II.17)
\Phi(\mathbb{P},t)=\frac{1}{\tau_0}\int\limits_0^{t}
\frac{a(t')\beta_0(\mathbb{P},t')d
t'}{\sqrt{a^2(t')+\mathbb{P}^2/m_X^2}}.
\end{equation}
According to the formulae (28)-(35)\cite{Yu_barII} we come from
the temporal variable $t$ and the impulse one $\mathbb{P}$ to the
dimentionless temporal variable $\eta$ and the impulse variable
$\xi$
\begin{equation}\label{eta}
t=\tau_0 \eta.
\end{equation}
In doing so we choose the normalization of scale factor for a
completely ultrarelativistic stage of universe ex\-pan\-si\-on,
such one that
\begin{equation}\label{a(eta)}
a(t)=\sqrt{\tau_0\eta},
\end{equation}
where $\tau_0$ is the supermassive boson decay time in its own frame of reference
\begin{equation} \label{III.123}%(II.18)
\tau_0=\frac{4\pi m_X}{s^2} \sim \frac{3}{2}(m_X \alpha)^{-1}.
\end{equation}
and introduce a dimentionless parameter $\sigma$
\begin{equation}\label{III.129}
\sigma =
\frac{m_X}{T(\tau_0)}=\frac{m_X\sqrt{\tau_0}}{\mathcal{T}_0} =
\frac{\chi \sqrt{m_X}}{\sqrt{\alpha} \mathcal{T}_0},
\end{equation}
The dimentionless impulse variable $\xi$ let us introduce with the
help of the relation (Ref. \cite{Yu_barII}):
\begin{equation}\label{xi}
\mathbb{P}=m_x\sqrt{\tau_0}\xiá \Rightarrow
\xi=\frac{\mathbb{P}}{m_X\sqrt{\tau_0}} \end{equation}
so that
\begin{equation}\label{E/T}
\frac{E}{T}=\sqrt{m^2_X+\mathbb{P}^2/a^2(t)}{T}=\sigma\sqrt{\eta+\xi^2}
\end{equation}
and the equilibrium function of the supermassive bosons is equal
to (Ref. (13)\cite{Yu_barII})
\begin{equation}\label{f_0}
f_0(\eta,\xi)=\frac{1}{e^{\sigma\sqrt{\eta+\xi^2}}-1}.
\end{equation}
As the done investigations have shown the final results of the
investigations are very weakly sensitive to statistic factors
consideration by the factors $\Phi(\eta,\xi)$ calculating, whereas
they are very sensitive to the static factors consideration at the
other stages of calculation. Therefore further on we shall
calculate the function $\eta,\xi$ in the Boltzmann approximation ,
while at the other stages of calculation we shall hold the
statistic factors. Then in the Boltzmann approximation
$\beta_0\approx 1/2$, and we obtain for the function $\Phi$:
\begin{equation} \label{dotPhi}
\dot{\Phi}(\mathbb{P},t)=\frac{1}{2\tau_0}\frac{\sqrt{\eta}}{\sqrt{\eta+\xi^2}}
\end{equation}
and -
\begin{equation} \label{Phi}
\Phi(\xi,\eta)=\frac{1}{2}\left(\sqrt{\eta}\sqrt{\eta+\xi^2}-
\xi^2\ln \frac{\sqrt{\eta}+\sqrt{\eta+\xi^2}}{\xi}\right).
\end{equation}
In consequence of (\ref{dotPhi}) $G(\eta,\xi)$ is a nonnegative
mo\-no\-to\-ni\-cal\-ly increasing function
\begin{equation}\label{Phi>0}
\Phi(\eta,\xi)\geq 0; \quad \frac{\partial
\Phi(\eta,\xi)}{\partial
\eta}=\frac{1}{2}\frac{\sqrt{\eta}}{\sqrt{\eta+\xi^2}}\geq 0,
\end{equation}
with
\begin{equation}\label{Phi=0}
\lim\limits_{\eta\to 0}\Phi(\eta,\xi)=0.
\end{equation}
Note that the rather large expression for the function
$\Phi(\xi,\eta)$, obtained in \cite{Yu_barII}, can be reduced to
the given above after simple transformations.

Thus coming to the new variables we get an expression for specific
entropy

\begin{equation} \label{ds/N_eta}
\delta_S=\frac{15 \Delta r
\mathcal{N}_X}{2\pi^6\mathcal{N}}\sigma^3\int\limits_0^\infty
d\eta e^{-\Theta(\eta)}\sqrt{\eta}\int\limits_0^\infty
\frac{\xi^2d\xi}{\sqrt{\eta+\xi^2}}\delta f(\eta,\xi),
\end{equation}
where
$$\Theta(\eta)=\int\limits_t^\infty \Psi(t)dt.$$
As far as $\dot{\Phi}\geq 0$, then $\Psi(t)>0$, thus
$\Theta(\eta)$ is a nonnegative monotonically decreasing function
\begin{equation}\label{theta'}
\frac{d \Theta}{d\eta}\leq 0.
\end{equation}
In this case

\begin{equation}\label{df}
\delta f(\eta,\xi)=e^{-\Phi(\eta,\xi)}\int\limits_0^\eta d\eta'
e^{\Phi(\eta',\xi)}\frac{\partial}{\partial
\eta'}\frac{1}{e^{\sigma\sqrt{\eta'+\xi^2}}-1}.
\end{equation}

\section{Functions $\Psi(t)$ and $\Theta(t)$}
Coming to the new variables in Boltzmann approximation of the
function $\Phi(x)$ (\ref{Phi}) the expression for the function
$\Psi(\eta)$, in which the statistic factor is already taken into consideration, is
\begin{equation}\label{Psi1}
\Psi(\eta)=\frac{\sqrt{\eta}N_X \sigma^3}{2\pi^2\tau_0}
\int\limits_0^\infty
{\displaystyle\frac{1}{e^{\sigma\sqrt{\eta+\xi^2}}-1}}
\frac{\xi^2d\xi}{\sqrt{\eta+\xi^2}}\end{equation}
Let us introduce new variables $x$ and $z$
\begin{equation}\label{x&z}
\xi=\sqrt{\eta}\sinh(x),\quad z=\sigma\sqrt{\eta}.\end{equation}
Then we get
$$\Psi(\eta)=\frac{N_X}{\pi^2\tau_0}z^3\int\limits_0^\infty
{\displaystyle \frac{\sinh^2 t dt}{e^{z\cosh t}-1}}.$$

Calculating the integral $\int\Psi dt$ and changing the order
of integration in the obtained expression we arrive at
\begin{equation}\label{Theta(eta)}
\Theta(\eta)=\int\limits_t^\infty\Psi(t')dt'=\frac{2N_X}{\pi^2\sigma^2}
\int\limits_0^\infty\frac{\sinh^2 x}{\cosh^5
x}dx\int\limits_{\sqrt{\eta}\sigma\cosh
x}^\infty\frac{\nu^3}{e^\nu -1}d\nu.\end{equation}
In particular integrating over the whole interval of the value
$t$ we obtain the integrals product
$$\Theta(0)=
\int\limits_0^\infty\Psi(t')dt'=\frac{2N_X}{\pi^2\sigma^2}
\int\limits_0^\infty\frac{\sinh^2 x}{\cosh^5
x}dx\int\limits_0^\infty\frac{\nu^3}{e^\nu -1}d\nu,$$ one of them
is expressed in terms of $\zeta$ - Riemann function
$$\int\limits_0^\infty \frac{\nu^3}{e^\nu -1}=\frac{\pi^4}{15},$$
and the other is easily calculated
$$\int\limits_0^\infty \frac{\sinh^2x}{\cosh^5x}dx=\frac{\pi}{16}.$$
As a result we find
\begin{equation}\Theta(0)
=\int\limits_0^\infty\Psi(t')dt'=\frac{\pi^3N_X}{120\sigma^2}.\end{equation}
Thus we can write down
$$\Theta(\eta)=\Theta(0)-\frac{2N_X}{\pi^2\sigma^2}
\int\limits_0^\infty\frac{\sinh^2 x}{\cosh^5
x}dx\int\limits_0^{\sqrt{\eta}\sigma\cosh x}\frac{\nu^3}{e^\nu
-1}d\nu.$$
The inner integral can be expressed in terms of the
function\footnote{The function $D(x)$ is connected with the Debye
functions (Ref., for example \cite{Janke}).}
$$D(x)=\frac{3}{x^3}\int\limits_0^x \frac{t^3}{e^t-1}dt.$$
the function $D(x)$ has the following asymptotics
\begin{equation}\label{D_ap}
D(x)\approx\left\{\begin{array}{ll}
{\displaystyle 3\sum\limits_0^\infty \frac{B_n}{(n+3)n!}\,x^n }, & x\lesssim 1;\\
  & \\
{\displaystyle \frac{\pi^4}{5x^3}
-3\left(1+\frac{3}{x}+\frac{6}{x^2}\right)e^{-x}}, & x\gg 1,\\
\end{array} \right.
\end{equation}
where $B_n$ are Bernoulli numbers. The function $D(X)$ graph is
shown in Fig. \ref{DG}.
\vskip 24pt\noindent \refstepcounter{figure}
\centerline{\epsfig{file=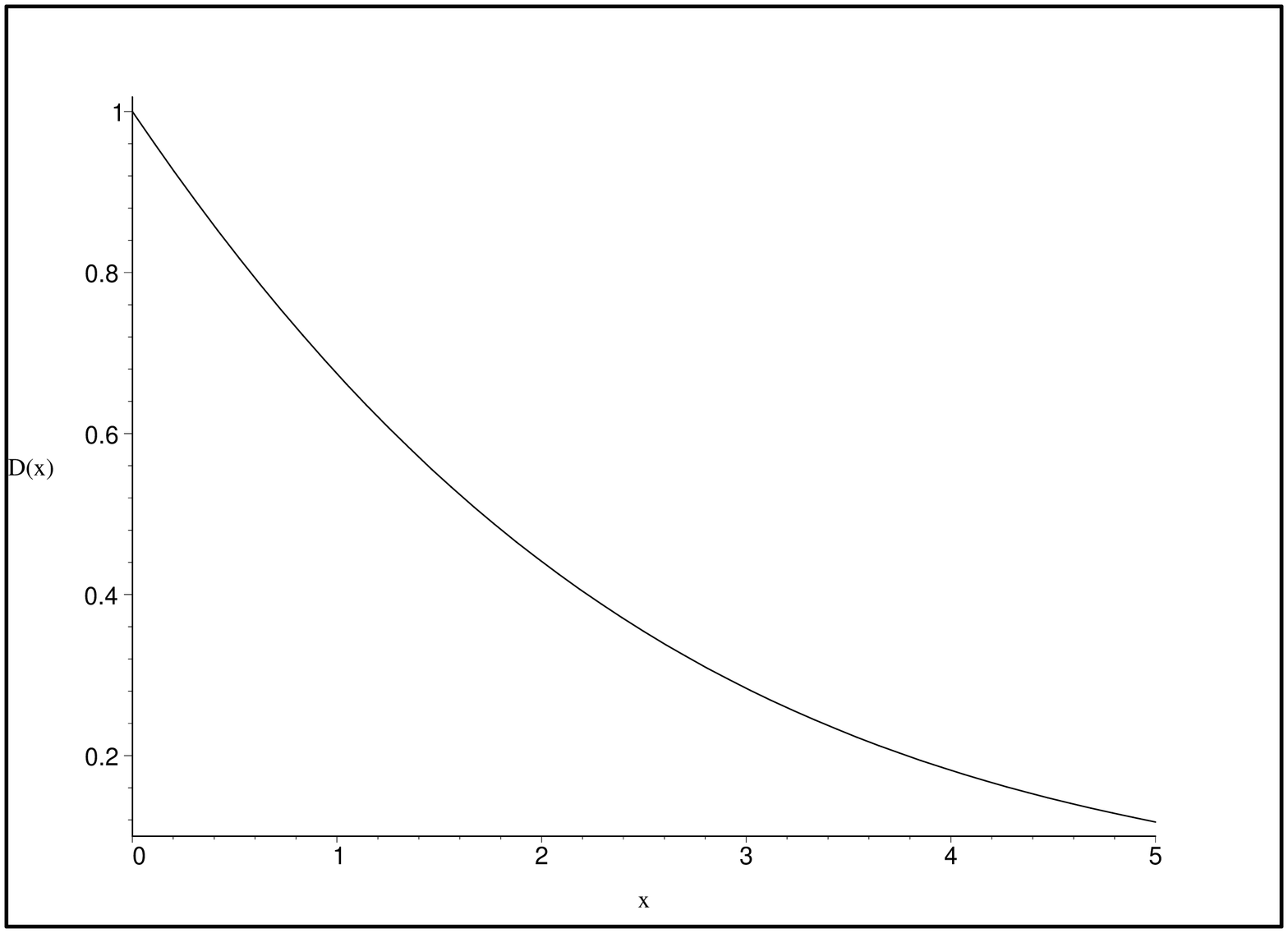,height=8cm,width=8cm,angle=-90}}\label{DG}
\vskip 12pt \noindent {Fig.\bf \thefigure.}\hskip 12pt{\sl The
function $D(x)$ graph constructed with the help of the Maple
package on the approximations of the type (\ref{D_ap}) .\hfill}
\vskip 12pt\noindent%
Thus finally we obtain
\begin{equation}\label{Theta(eta)1}
\Theta(\eta)= \frac{\pi^3
N_X}{120\sigma^2}\Xi(\sqrt{\eta}\sigma),\end{equation}
where a monotonically decreasing function $\Xi(x)$ is introduced
\begin{equation} \label{Xi}
\Xi(x)=1-\frac{80}{\pi^5}x^3\int\limits_0^\infty \tanh^2x D(x\cosh
z)dz,\end{equation}
varying in the interval $$ 0\geq \Xi(x)\leq 1.$$
To calculate with the functions $D(x)$ and $\Xi(x)$ a special
library in the package of symbol Maple Mathematics was worked out
in which the proceeders of time-optimal calculation of these
functions with the help of different approximations are
determined. The function $\Xi(x)$ graph is shown in Fig.
\ref{Xi(x)}
\vskip 24pt\noindent \refstepcounter{figure}
\centerline{\epsfig{file=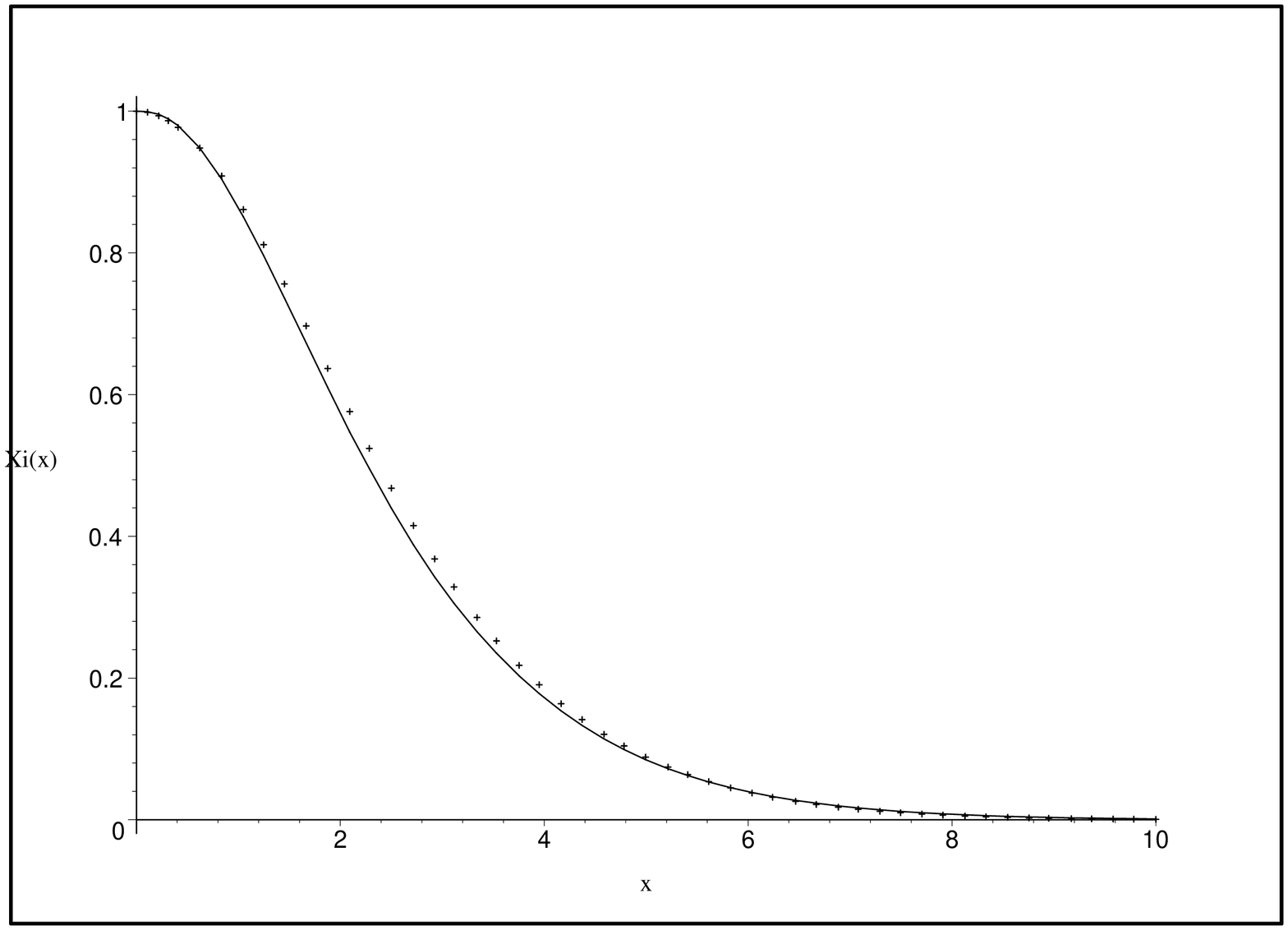,height=7cm,width=7cm,angle=-90}}\vskip
12pt \noindent {Fig.\bf \thefigure.}\hskip 12pt {\sl The function
$\Xi(x)$ graph. The dotted line denotes the extrapolating function
$F(x)$ graph (\ref{extrapol}). \label{Xi(x)}}\hfill
\vskip 12pt\noindent%
In the interval $[0,10]$ the function $\Xi(x)$ is well
extrapolated by the function
\begin{equation}\label{extrapol}
F(x)=\frac{e^{-\alpha x^2}}{1+\beta x^2}\end{equation}
with the parameters $\alpha=0,05$ and $\beta=0,09$.

\section{Equilibrium deviation $\delta f(\eta,\xi)$}
Now let us calculate the function $\delta f(\eta,\xi)$. As it is
not difficult to see the function $\Phi(\eta,\xi)$ is a slowly
varying function, because in consequence (\ref{Phi>0})
$$\Phi'_\eta <1/2,$$
while
$$\lim\limits_{\eta\to 0}e^{\Phi(\eta,\xi)}=1,$$
and in wide limits of changing the variables $\eta,\xi$:
$$\exp(\Phi(\eta,\xi))\approx 1.$$ And the derivative of the
equilibrium  function of dis\-t\-ri\-bu\-ti\-on has the order
\begin{equation}\label{dot_f0}
\frac{f'_0}{f_0}=-\frac{\sigma}{2\sqrt{\eta
+\xi^2}}\frac{e^{\sigma\sqrt{\eta+\xi^2}}}{e^{\sigma\sqrt{\eta+\xi^2}}-1}
\end{equation}
and infinitely grows in the range $\sigma\sqrt{\eta+\xi^2}\to 0$;
By large values of this argument this variable is small. Thus,
values of the variables $\eta,\xi$ in the range
$\sqrt{\eta+\xi^2}\lesssim \sigma^{-1}$ in which $\exp(\Phi)$ can
be considered approximately constant make the main contribution to
the distribution function deviation from equilibrium. Thus,
integrating by parts in (\ref{df}) we find approximately
\begin{equation}\label{df_appr}
\delta f(\eta,\xi) \simeq \frac{1}{e^{\sigma
\xi}-1}-\frac{e^{-\Phi(\eta,\xi)}}{e^{\sigma\sqrt{\eta+\xi^2}}-1}.
\end{equation}
Substituting this function into the inner integral
(\ref{ds/N_eta}) and introducing the function
\begin{equation}\label{Df2}
{\rm Df}(\eta,\sigma)=\sigma^2\int\limits_0^\infty \frac{\xi^2
d\xi}{\sqrt{\eta+\xi^2}}\left[\frac{1}{e^{\sigma
\xi}-1}-\frac{e^{-\Phi(\eta,\xi)}}{e^{\sigma\sqrt{\eta+\xi^2}}-1}\right]
\end{equation}
we obtain
\begin{equation} \label{ds_Df}
\delta_S=\frac{15 \Delta r
\mathcal{N}_X}{2\pi^6\mathcal{N}}\sigma\int\limits_0^\infty d\eta
e^{-\Theta(\eta)}\sqrt{\eta}{\rm Df}(\eta,\sigma).
\end{equation}
Note, the introduced earlier function ${\rm Df}(\eta,\sigma)$ is
pro\-por\-ti\-o\-nal to the perturbed track of tensor of  $X$ -
bosons energy-impulse
$$\delta T_X = g_{ik}\delta T^{ik}_X=m^2_X\int\limits_{P(X)}\delta f_X dP.$$
Coming to the numerical integration in (\ref{ds_Df}) let us note
that the inconvenient for a numerical integration integrals are
just in in the function ${\rm Df}(\eta,\sigma)$, so the direct
application of numerical integration runs against the divergence
problem. Therefore, at first the integral (\ref{Df2}) is necessary
to be transformed into a  convenient for a numerical integration
type. For this let us rewrite the integral (\ref{Df2}) in the
equivalent form
$$\!\!{\rm Df}(\eta,\sigma)=\sigma^2\!\left[\int\limits_0^\infty \frac{\xi^2d\xi}{\sqrt{\eta+\xi^2}}
\left(\frac{1}{e^{\sigma\xi}-1}-\frac{1}{e^{\sigma\sqrt{\eta+\xi^2}}-1}
\right)\right.\!+$$
$$\left.+\int\limits_0^\infty
\frac{\xi^2d\xi}{\sqrt{\eta+\xi^2}}{\displaystyle
\frac{1-e^{-\Phi(\eta,\xi)}}{e^{\sigma\sqrt{\eta+\xi^2}}-1}}
\right].$$
Let us study the first part of the integral
$$A=\int\limits_0^\infty \frac{\xi^2d\xi}{\sqrt{\eta+\xi^2}}
\left(\frac{1}{e^{\sigma\xi}-1}-\frac{1}{e^{\sigma\sqrt{\eta+\xi^2}}-1}
\right)\equiv$$
$$\int\limits_0^\infty \frac{\xi^2d\xi}{\sqrt{\eta+\xi^2}}{\displaystyle
\frac{1}{e^{\sigma\xi}-1}}-\int\limits_0^\infty
\frac{\xi^2d\xi}{\sqrt{\eta+\xi^2}}{\displaystyle
\frac{1}{e^{\sigma\sqrt{\eta+\xi^2}}-1}}.
$$
in the first integral we make a substitution $\xi=\sqrt{\eta}x$,
and in the second $\xi=\sqrt{x^2-1}$. Then we get
$$\!A=\!\eta\left[\int\limits_0^\infty
\frac{x^2dx}{\sqrt{1+x^2}}{\displaystyle \frac{1}{e^{\sigma\sqrt{\eta}x}-1}}-
\int\limits_1^\infty {\displaystyle
\frac{\sqrt{\eta+x^2}dx}{e^{\sigma\sqrt{\eta}x}-1}}\right]\!.
$$
Bringing these integrals together we obtain
$$
\!A=\!\eta\left[\int\limits_0^1
\frac{x^2dx}{\sqrt{1+x^2}}{\displaystyle
\frac{1}{e^{\sigma\sqrt{\eta}x}-1}}+\int\limits_1^\infty
{\displaystyle
\frac{dx}{\sqrt{\eta+x^2}}\frac{1}{e^{\sigma\sqrt{\eta}x}-1}}
\right]\!.
$$
Now we transform the $B$ part of the integral
$$B=\int\limits_0^\infty
\frac{\xi^2d\xi}{\sqrt{\eta+\xi^2}}{\displaystyle
\frac{1-e^{-\Phi(\eta,\xi)}}{e^{\sigma\sqrt{\eta+\xi^2}}-1}}.$$
Substituting the expression for $\Phi(\eta,\xi)$ into the integral
and making a substitution $\xi=\sqrt{\eta}x$, we transform this
integral
$$
B=\eta\int\limits_0^\infty \frac{x^2dx}{\sqrt{1+x^2}}
{\displaystyle \frac{{\displaystyle
1-e^{-\frac{1}{2}\eta\left(\sqrt{1+x^2}-
x^2\ln\frac{1+\sqrt{1+x^2}}{x}\right)}}}{e^{\sigma\sqrt{\eta}\sqrt{1+x^2}}-1}}.
$$
In this integral the difficulties of numerical integration arise
under the conditions of $\sigma\sqrt{\eta}\to 0$. Expanding the
exponent into Tailor series small values $\eta$, we get
approximately
$$
B\approx \frac{1}{2}\eta^2\int\limits_0^\infty
\frac{x^2dx}{\sqrt{1+x^2}} {\displaystyle
\frac{\sqrt{1+x^2}-x^2\ln\frac{1+\sqrt{1+x^2}}{x}}
{e^{\sigma\sqrt{\eta}\sqrt{1+x^2}}-1}}.
$$
Numerical integration of these expression does not run to any
difficulties already. Considering the above comments in the Maple
package there was created a library of special procedures of rapid
calculation of the function ${\rm Df}(\eta,\sigma)$ for any values
of variables. The graphs of the functions ${\rm Df}(\eta,\sigma)$
obtained with the
help of these procedures are shown in Fig. \ref{GDf}. %
\vskip 24pt\noindent \refstepcounter{figure}
\centerline{\epsfig{file=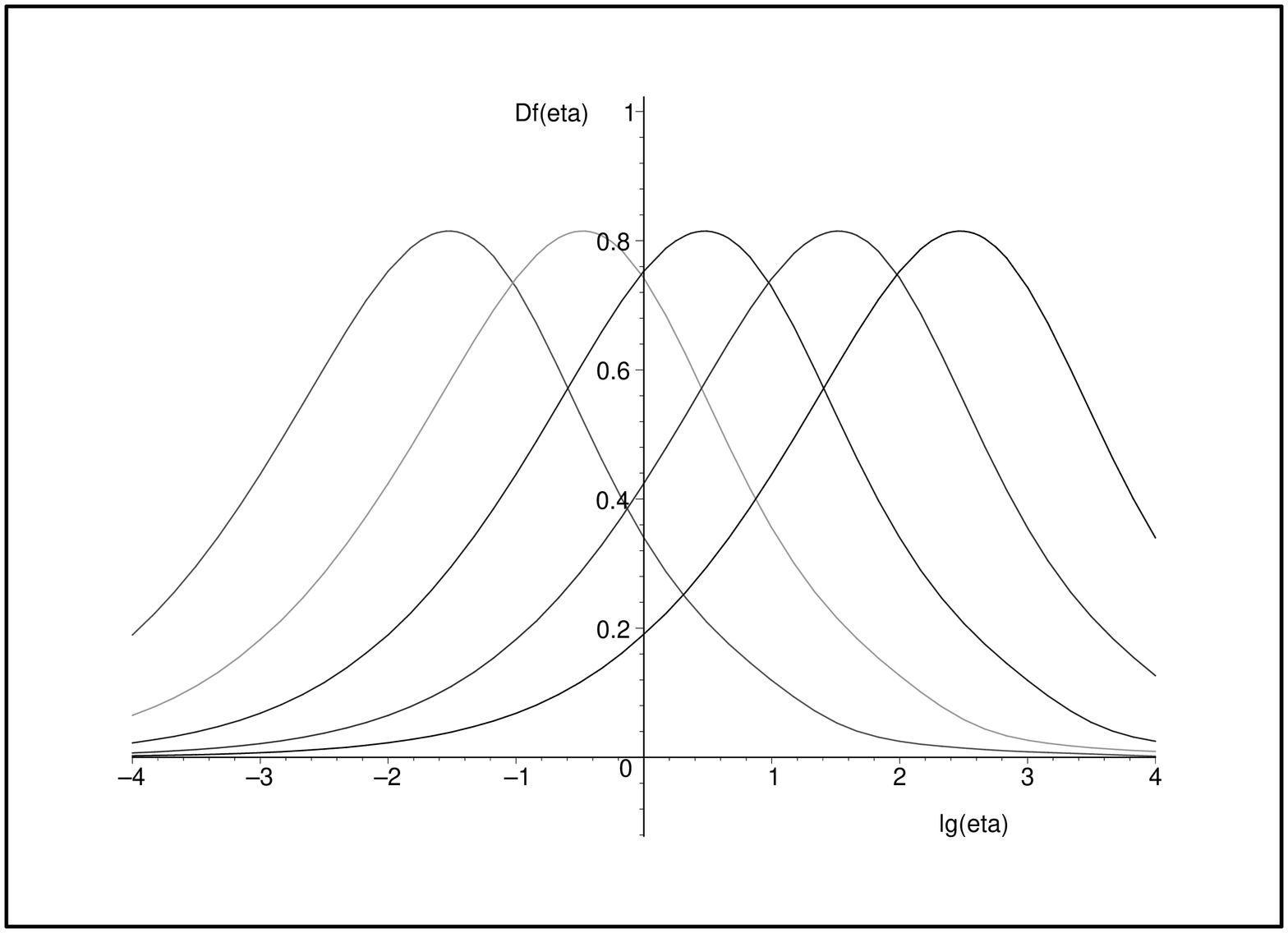,height=7cm,width=7cm,angle=-90}}\vskip
12pt \noindent {Fig.\bf \thefigure.}\hskip 12pt{\sl Functions
${\rm Df}(\eta,\sigma)$ subjecting to the parameter $\sigma$.
Along the abscissa axis $\lg \eta$ is plotted. From the left to
the right $\sigma=10$, $\sigma=3$, $\sigma=1$, $\sigma=0,3$,
$\sigma=0,1$. \label{GDf} }.\hfill
\vskip 12pt\noindent%

\section{Results}
Before proceeding to the results presentation let us carry out a
convenient function $\delta_S$ normalization.  As noted in
\cite{Yu_barI}, in the papers \cite{OtheIgnat}-\cite{Weinberg1}
the specific entropy estimation for one baryon was obtained
(formula (5) \cite{Yu_barI}):
\begin{equation}\label{delta_s0}
\delta^0_S=\frac{45\zeta(3)}{4\pi^4}\frac{N_X}{N}\Delta r .
\end{equation}
Therefore we shall correspond our results to this estimation by introducing
a relative variable
\begin{equation}\label{Delta_s}
\Delta_S=\frac{\delta_S}{\delta_S^0}
\end{equation}
so called the reduced specific entropy. Carrying out numerical
in\-teg\-ra\-ti\-on in the expression (\ref{ds_Df}) with the help
of the given procedures in the Maple package we obtain a graphs
family of the function $\Delta_S(\sigma)$. In Fig. \ref{ds} we
show calculated graphs of dependence $\Delta_S(\sigma)$ by
different values of $N_X$ - number of $X-$ boson types which is a
parameter of the model of interactions.
\vskip 24pt\noindent \refstepcounter{figure}
\centerline{\epsfig{file=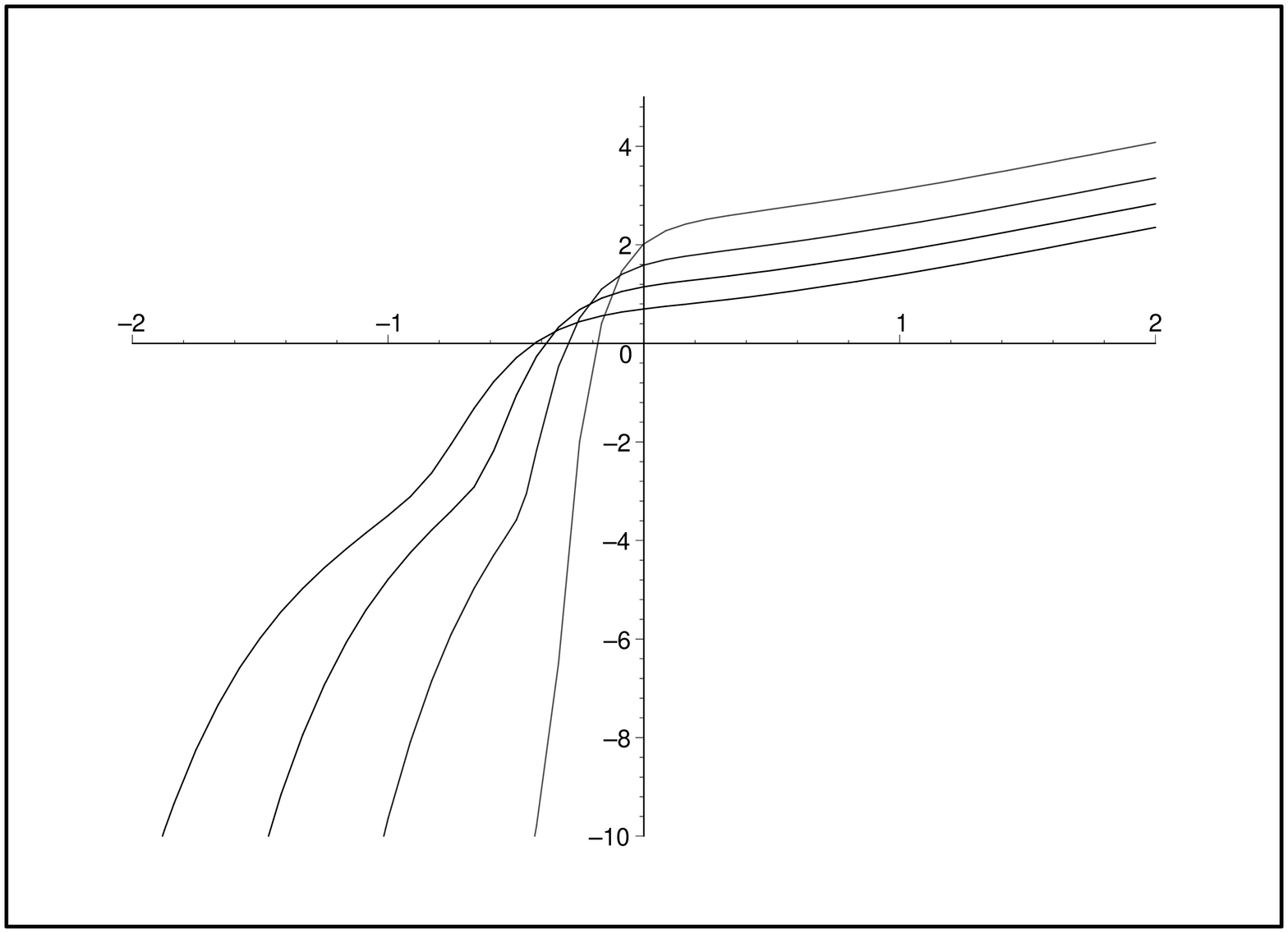,height=7cm,width=7cm,angle=-90}}\vskip
12pt \noindent {Fig.\bf \thefigure.}\hskip 12pt{\sl The given
specific entropy for one baryon, $\lg \Delta_S(\sigma,N_X)$,
subjecting to the number of $X$-bosons, $N_X$. Along abscissa axis
the values of $\lg \sigma$ are plotted. In the right part of the
figure from bottom up $N_X=1$; $N_X=3$; $N_X=10$; $N_X=53$.
\label{ds} }.\hfill
\vskip 12pt\noindent%

Passing on the analysis of the results we firstly note  that the
calculations carried out in terms of complete kinetic theory
Showed a sufficient dependence of the produced baryon charge in
the quantity of $X$ - bosons types. Let us point out the following
general tendency of this dependence of the produced baryon charge:
at $\sigma\lesssim0,4\div0,8$ with the increase of the number of
$X$ - bosons types the given specific entropy increases , and at
$\sigma\gtrsim0,4\div0,8$ on the contrary it decreases, moreover
in the range of small values of the parameter $\sigma$ the
dependence of the given entropy on $N_X$ is especially
perceptible. At the same time we should remember that the absolute
value of specific entropy equals to
\begin{equation}\label{delta_s}
\delta_S=\Delta_S\delta_S^0=\Delta_S\frac{45\zeta(3)}{4\pi^4}\frac{N_X}{N}
\Delta r.
\end{equation}
On the other hand we can suppose the factor $N_X/N$ (relation of
the number of $X$-bosons types to the general number of particles
types) does not strongly depend on the field model of
interactions, therefore the conclusion concerning dependence of
the number of $X$ - bosons types of the given entropy can be
carefully transferred to the absolute value of specific entropy
also. These peculiarities of dependence of specific entropy on the
number of $X$ - bosons types are shown in Fig. \ref{Ds_N}.
\vskip 24pt\noindent \refstepcounter{figure}
\centerline{\epsfig{file=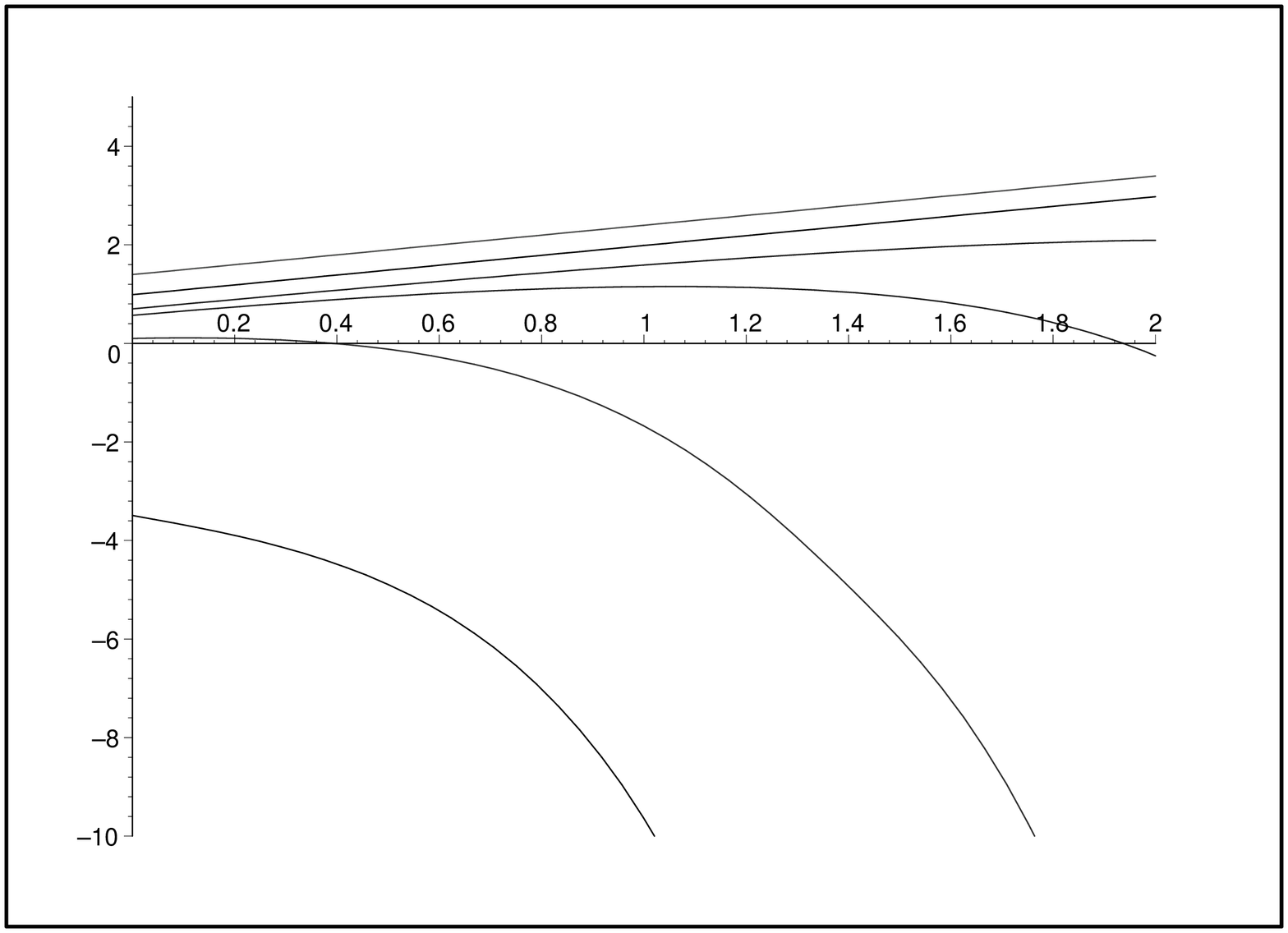,height=7cm,width=7cm,angle=-90}}\vskip
12pt \noindent {Fig.\bf \thefigure.}\hskip 12pt{\sl Dependence of
the given specific entropy for one baryon on the number of $X$ -
bosons types. Bottom-up $\sigma =0,1$; $\sigma=0,4$; $\sigma=0,7$;
$\sigma=1$; $\sigma=3$; $\sigma=10$. Along the abscissa axis the
values of $\lg N_X$ are plotted, along the ordinate axis the
values of $\lg \Delta_S$ are.\label{Ds_N}.}\hfill
\vskip 12pt\noindent%

Further, the value (\ref{delta_s0}) obtained by a number of
authors, in the kinetic theory is reached at the values of the
parameter $\sigma=0,4\div0,6$. Moreover the kinetic model of
cosmological baryogenesis detected a finer structure of this
process than the one which was produced by the hydrodynamic theory
of this process developed earlier in the quoted papers
\cite{Fry1}-\cite{Fry3}. The difference of our results from the
quoted papers results, especially in the range of small values of
the parameter $\sigma$, is caused by essential influence of
nonequilibrium processes on the final result in this range. It is
not difficult to see that it is this range where the function of
$X$-bosons distribution most differs from the equilibrium one. Of
course this fact cannot be taken into consideration in the
hydrodynamic model of bariogenesis effectively.

\end{document}